\documentclass[10pt]{article}

\usepackage{url,graphics}
\usepackage{multirow}
\usepackage{amsmath,amsthm}
\usepackage{algorithm}
\usepackage{algpseudocode}

\graphicspath{{/}}
\usepackage{amsmath}
\usepackage{amssymb}

\usepackage{graphicx}

\usepackage{cite}

\usepackage{color} 

\usepackage[labelfont=bf,labelsep=period,justification=raggedright]{caption}

\makeatletter
\renewcommand{\@biblabel}[1]{\quad#1.}
\makeatother

\date{}

\begin{document}

\begin{flushleft}
{\Large
\textbf{Inferring Disease and Gene Set Associations with Rank Coherence in Networks}
}
\\
TaeHyun Hwang$^{1}$, Wei Zhang$^{1}$,Maoqiang Xie$^2$,
Rui Kuang$^{1,\ast}$
\\
\bf{1} Department of Computer Science and Engineering, University of Minnesota Twin Cities, MN, USA
\\
\bf{2} College of Software, Nankai University, Tianjin, China\\
$\ast$ Correspondence: kuang@cs.umn.edu
\end{flushleft}

\section*{Abstract}
A computational challenge to validate the candidate disease genes identified in a high-throughput genomic study is to elucidate the associations between the set of candidate genes and disease phenotypes. The conventional gene set enrichment analysis often fails to reveal associations between disease phenotypes and the gene sets with a short list of poorly annotated genes, because the existing annotations of disease causative genes are incomplete. We propose a network-based computational approach called rcNet to discover the associations between gene sets and disease phenotypes.  Assuming  coherent associations between the genes ranked by their relevance to the query gene set, and the disease phenotypes ranked by their relevance to the hidden target disease phenotypes of the query gene set, we formulate a learning framework maximizing the rank coherence with respect to the known disease phenotype-gene associations. 
An efficient algorithm coupling ridge regression with label propagation, and two variants are introduced to find the optimal solution of the framework.
We evaluated the rcNet algorithms and existing baseline methods with both leave-one-out cross-validation and a task of predicting recently discovered disease-gene associations in OMIM. The experiments demonstrated that the rcNet algorithms achieved the best overall rankings compared to the baselines. To further validate the reproducibility of the performance, we applied the algorithms to identify the target diseases of novel candidate disease genes obtained from recent studies of GWAS, DNA copy number variation analysis, and gene expression profiling. The algorithms ranked the target disease of the candidate genes at the top of the rank list in many cases across all the three case studies. The rcNet algorithms are available as a webtool for disease and gene set association analysis at \url{http://compbio.cs.umn.edu/dgsa_rcNet}.
\section*{Author Summary}

\section*{Introduction}
Determination of the molecular cause of diseases is a major focus in genomics research since early 1960s \cite{Mc:omim}. Recently, powered by the advanced high-throughput genomic technologies, numerous large-scale genome-wide disease studies such as  genome-wide association studies \cite{GWAS1, Johnson09}, DNA copy number detections \cite{Shlien09}, and gene expression profiling \cite{Veer08}, were conducted towards this goal. Typically, the objective of a study is to perform a high-throughput scanning for a list of genes that are involved with the disease under study, and then a standard follow-up enrichment analysis or its variants and extensions is applied to analyze the gene set, based on the statistical significance of the overlap between the genes and gene functional annotations or associations with disease phenotypes. Examples of the well-known tools are DAVID \cite{HuangDA09}, GSEA \cite{GSEA05}, GOToolBox \cite{GOToolBox} and many others. However, in many cases, since the existing annotations of disease causative genes is far from complete \cite{Mc:omim}, and a gene set might only contain a short list of poorly annotated genes, enrichment-based approaches often fail to reveal the associations between gene sets and disease phenotypes. 

The availability of large phenotypic and molecular networks provides a new opportunity to study the association between diseases and the gene sets identified from the high-throughput genomic studies. The human disease phenotype network \cite{van:text} provides information on phenotype similarities computed by text mining of the full text and clinical synopsis of the disease phenotypes in OMIM \cite{Mc:omim}. Large molecular networks such as the human protein-protein interaction network \cite{wu:network} or functional linkage network \cite{Linghu09} provide functional relations among genes or proteins. Based on the observation that genes associated with the same or related diseases tend to interact with each other in the gene network, many network-based approaches are proposed to utilize the disease modules and gene modules in the networks to prioritize disease genes, a task of ranking genes for studying genetic diseases \cite{franke:reconstruction, sebastian08, wu:network, Linghu09, Hwang:heterogeneous, vanunu:assocating, Li:genome-wide}.    

\begin{figure*}
  \centering
  \includegraphics[scale=0.50]{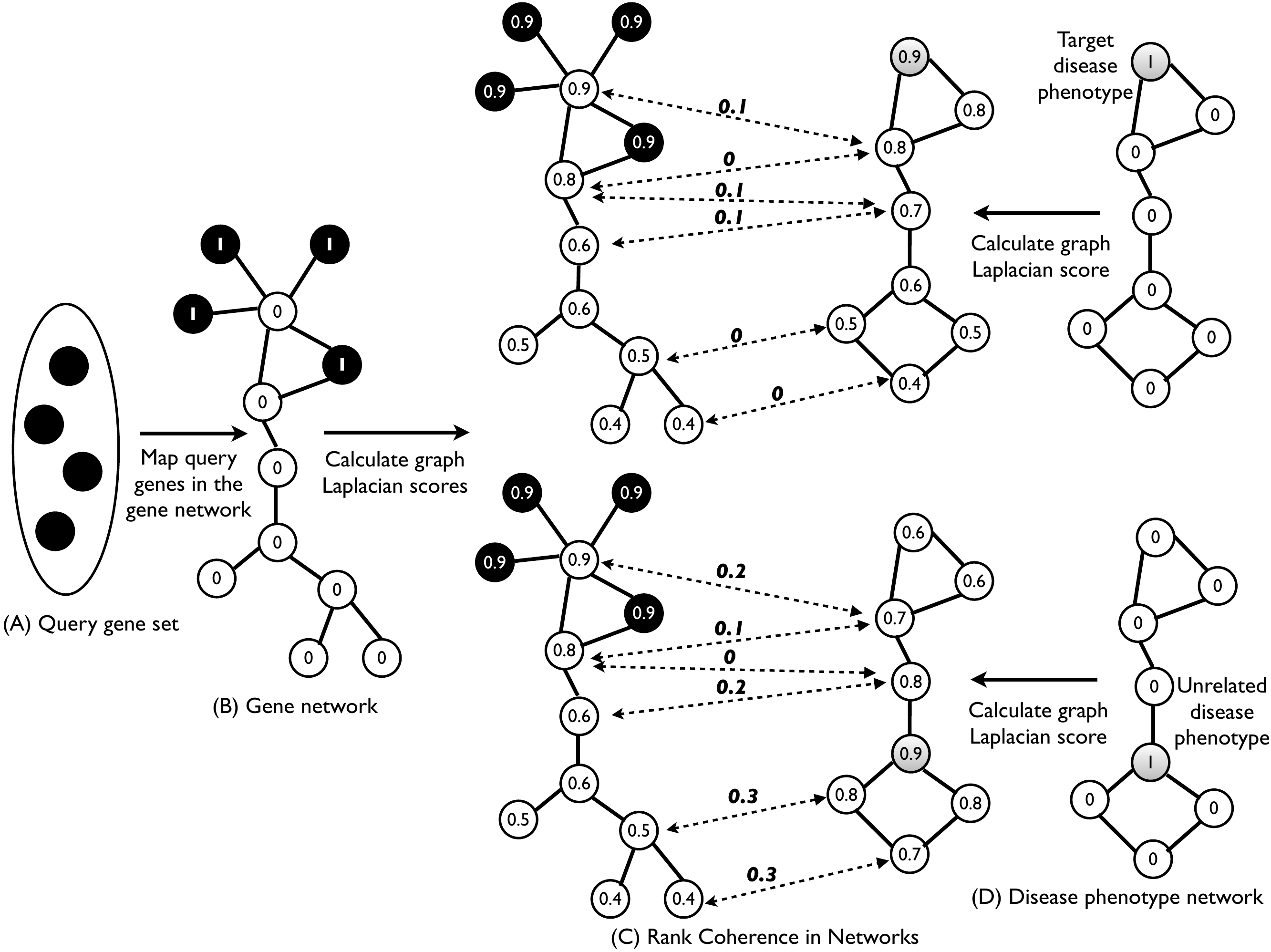}
  \vspace{-10pt}
\caption{{\bf Illustration of Rank Coherence in Networks.}  A query gene set of four genes is given in {\bf (A)}. The four genes are mapped in the gene network and the corresponding nodes are marked with 1 in {\bf (B)}. The graph Laplacian scores are then computed to quantify the relevance between each gene (including the query genes) and the query gene set. Similarly, if a disease phenotype of the gene set is selected and marked with 1, the graph Laplacian scores can be derived to quantify the relevance between each disease phenotype and the selected phenotype in {\bf (D)}. Based on the coherence assumption, the top-ranked genes and the top-ranked phenotypes should be highly connected with each other if the phenotype is the target of the query gene set, otherwise the connectivity will be close to random. As showed in {\bf (C)}, the edges connecting associated genes and phenotypes are labeled by the discrepancy between their ranking scores. Clearly, the phenotype ranking given by target phenotype query is more coherent (the upper case) than the ranking given by an unrelated phenotype (the bottom case). The connectivity is measured by Rank Coherence in the Networks (rcNet). In general, since the target disease phenotypes are not known, the rcNet algorithms search for the phenotype with the best rcNet score against the query gene set. } \label{fig:framework}
  \vspace{-10pt}
\end{figure*}

In this paper, we propose a general network-based approach to infer associations between disease phenotypes and gene sets, utilizing the disease phenotype network and the gene network. We formulate the problem as a gene set query problem. By querying the networks with a given gene set, a user expects to retrieve a list of disease phenotypes with the highest predicted association with the gene set. Based on the assumption that the genes ranked by their relevance to the query gene set will have coherent associations with the disease phenotypes ranked by their relevance to the hidden target disease phenotypes, we formulate a simple learning framework maximizing Rank Coherence in Networks (rcNet) with respect to the known disease phenotype-gene associations in OMIM. Fig. \ref{fig:framework} illustrates the general idea of Rank Coherence in Networks. We first measure the global relevance between the query gene set and all the genes with graph Laplacian scores (Fig. \ref{fig:framework}A\&B). The Laplacian scores can be considered as the result of using the query gene set as the seed to perform random walk with restart (or label propagation) in the gene network \cite{bengio:label}. The global relevance between a target disease phenotype and all disease phenotypes can be similarly computed as the Laplacian scores with random walk on the disease phenotype network (Fig. \ref{fig:framework}D). Our assumption is that, between the rankings given by the query gene set and the target disease phenotype, the top-ranked genes and the top-ranked phenotypes should be highly connected by known associations, quantified by Rank Coherence in Networks (Fig. \ref{fig:framework}C). In a real problem, the target disease phenotypes are unknown. The rcNet algorithms are designed to search for the phenotype(s) with the best rcNet score against the query gene set. 
We propose two strategies. The first approach relaxes the combinatorial problem as ridge regression to find a closed-form solution for selecting the target disease phenotype. The second approach in two variants enumerates all possible phenotype configurations to find the best match of the query gene set.


The rcNet algorithms are different from the gene set enrichment analysis with statistical methods such as Hypergeometric statistics, permutation test or non-parametric McNemar's test \cite{HuangDA09, GSEA05, GOToolBox} because the rcNet algorithms use the topological information in the disease phenotype network and the gene network to analyze the association between a gene set and all phenotypes simultaneously. The simultaneous analysis of all phenotypes provides a global dependence, and thus richer and more reliable information for computing the association scores are used to rank the phenotypes. The rcNet algorithms share more algorithmic similarity with the disease gene prioritization methods, which were proposed for a different purpose. CIPHER \cite{wu:network} scores each gene against a disease phenotype based on the correlation between their relevances with all the phenotypes, where the relevance between the gene and a phenotype is calculated based on the distance between the gene and the genes associated with the phenotype. The methods proposed by \cite{sebastian08}, \cite{vanunu:assocating} and \cite{Linghu09} applied random walk (label propagation) or simpler neighborhood weighting to exploit  the gene networks for ranking genes for a disease phenotype, based on the seed genes mapped from the disease phenotype. One limitation is that the phenotype network and the sparse known associations are not fully utilized in the global analysis. The label propagation algorithms proposed by \cite{Hwang:heterogeneous} and \cite{Li:genome-wide} explore a heterogeneous network combining the gene  network, the phenotype network and the associations to explore gene modules, phenotype modules and the phenotype-gene association biclusters. Since the two methods make full use of the information in the networks, it is difficult to interpret the results and to tune the best parameters for combining the information.   

\section*{Methods}
\begin{figure} 
\centering
\framebox{
  \begin{minipage}{0.45\textwidth}
    $\mathbf{dgsa\_rcNet(g, \bar{G}, \bar{P}, A,\alpha,\beta)}$
    \begin{enumerate}
    \item[1] $\mathbf{p = 0}$ 
    \item[2] $\mathbf{\tilde{g}=(1-\alpha)(I-\alpha \bar{G})^{-1}g}$ (equation \eqref{eqn:alg}).
    \item[3] $\mathbf{\bar{A}=(1-\beta)A(I-\beta \bar{P})^{-1}}$
    \item[4] $\mathbf{p^* = (\bar{A}^T\bar{A} + \kappa I)^{-1}\bar{A}^T \tilde{g}}$
    \item[5] $\mathbf{p(p^*>a)=1}$  \textit{(target selection with threshold $\mathbf{a}$)}
    \item[6] \textbf{return} ($\mathbf{p}$)
    \end{enumerate}
  \end{minipage}
}
\caption{rcNet Algorithm - Rank Coherence in Networks.}\label{fig:rcNet_alg}
\end{figure}

\subsection*{Problem Definition}
We formulate a graph query problem for disease phenotype and gene set association discovery: given a heterogenous network consisting of the gene network, the phenotype network and the association network, we query the network with a gene set to retrieve a phenotype (or several) predicted to have association with the query gene set. 
We define $\mathbf {G_{(n\times n)}}$, $\mathbf {P_{(m\times m)}}$, and $\mathbf {A_{(n\times m)}}$ as the adjacency matrix of the gene network, the disease network, and the disease-gene association network, respectively, where $\mathbf{n}$ is the number of genes and $\mathbf{m}$ is the number of disease phenotypes in the networks. The query gene set is represented by a binary vector $\mathbf{g =[g_1, g_2,..., g_n]^T}$ denoting the gene membership against the gene set, i.e. each $\mathbf{g_i=1}$ if gene $\mathbf{i}$ is in the query gene set, otherwise $\mathbf{0}$. Similarly, the list of target phenotype(s) is given by another binary vector $\mathbf{p =[p_1, p_2,..., p_m]^T}$ and phenotype $\mathbf{j}$ is a target phenotype if $\mathbf{p_j=1}$. Our objective is to find the $\mathbf{p}$ that gives the best rank coherence with the query gene set $\mathbf{g}$.

\subsection*{Computing Graph Laplacian Scores}
To fully utilize network topological information, we compute the global relevance score between the query gene set $\mathbf{g}$ and all the genes based on the graph Laplacian of the gene network $\mathbf {G_{(n\times n)}}$. We first normalize $\mathbf {G}$ as $\mathbf{\bar{G}=D_G^{\frac{1}{2}}GD_G^{\frac{1}{2}}}$, where $\mathbf{D_G}$ is a diagonal matrix with diagonal elements $\mathbf{D_{G i,i}=\sum_j G_{i,j}}$. A vector $\mathbf{\tilde{g}}$ of graph Laplacian scores is derived from the following optimization problem \cite{zhou:ranking},  
\begin{equation}
\mathbf{min_{\tilde{g}} \sum_{i,j} \bar{G}_{i,j}(\tilde{g}_i - \tilde{g}_j)^2 + \frac{1 - \alpha}{\alpha} \sum_i (\tilde{g}_i - g_i)^2} \label{eqn: genelap}
\end{equation}
In equation \eqref{eqn: genelap}, the first term is a smoothness penalty, which forces connected genes to receive similar scores, and the second term ensures the consistency with the query gene set. The Laplacian scores combine the neighboring information in the network with the consistency with the query gene set to provide a global relevance measure between each gene and the query gene set. Parameter $\mathbf{\alpha \in (0,1)}$ balances the contributions from the two penalties. The closed-form solution of equation \eqref{eqn: genelap} is 
\begin{equation} \label{eqn:gsol}
\mathbf{\tilde{g}=(1-\alpha)(I-\alpha \bar{G})^{-1}g}.
\end{equation}
Empirically, to avoid computing the inverse of $\mathbf{(I-\alpha \bar{G})}$,  an iterative algorithm can efficiently compute the closed-form solution with the following update rule at each time step $t$,
\begin{equation} 
\mathbf{\tilde{g}^t=(1-\alpha)g+\alpha \bar{G} \tilde{g}^{t-1}}, \label{eqn:alg}
\end{equation}
Similarly, graph Laplacian scores can be derived to measure the relevance between the phenotypes and the target phenotypes $\mathbf{p}$ with optimization of \begin{equation}
\mathbf{min_{\tilde{p}} \sum_{i,j} \bar{P}_{i,j}(\tilde{p}_i - \tilde{p}_j)^2 + \frac{1 - \beta}{\beta} \sum_i (\tilde{p}_i - p_i)^2} \label{eqn: phelap},
\end{equation}
with the closed-form solution 
\begin{equation}
\mathbf{\tilde{p}=(1-\beta)(I-\beta \bar{P})^{-1}p}, \label{eqn:psol}
\end{equation}
where $\mathbf{\bar{P}}$ is the normalized $\mathbf{P}$ and $\mathbf{\beta \in (0,1)}$ is the balancing parameter. Computing the laplacian scores is equivalent to a weighted summation of performing random walk on the graph with all the steps to infinite. Thus, the laplacian scores exploit modular information in a network to capture long range interactions between the nodes in a graph. Note that one can use other scoring functions such as counting the direct neighbors of the query gene set, or measuring the shortest path from the query gene set to other genes as suggested in \cite{wu:network}. However, empirically, the direct-neighbor function tends to generate very sparse information, and the shortest-path function does not fully explore the neighborhood information.

\subsection*{Rank Coherence in Networks}
Rank Coherence in Networks (rcNet) measures whether the query gene set $\mathbf{g}$ and a phenotype set $\mathbf{p}$ show coherent associations with the known disease-gene associations. Specifically, given the graph Laplacian scores $\mathbf{\tilde{g}}$, which rank the genes by their relevance to the query gene set $\mathbf{g}$, and  the graph Laplacian scores $\mathbf{\tilde{p}}$, which rank the disease phenotypes by their relevance to the hidden target phenotypes $\mathbf{p}$, Rank Coherence in Networks $\mathbf{rcNet(\tilde{g}, \tilde{p}, A)}$ measures whether the associations given by $A$ are connecting genes and phenotypes with similar scores in $\mathbf{\tilde{g}}$ and $\mathbf{\tilde{p}}$. We propose two different approaches to define Rank Coherence in Networks. The first approach adopts a ridge regression model coupled with label propagations to compute a closed-form solution of $\mathbf{p}$, relaxed to real numbers. The second approach uses simpler measures and enumerate all possible $\mathbf{p}$ to find the best fitting for $\mathbf{g}$.

\subsubsection*{A Ridge Regression Model}
Under the assumption that the Laplacian score of a phenotype can be reconstructed by the linear combination of the Laplacian scores of its gene neighbors in $A$, we can formulate the following least-square cost function,
\begin{equation} \label{eqn:reg}
\mathbf{ \Omega = || A \tilde{p} - \tilde{g}||^2}.
\end{equation}
Eventually, we are interested in deriving $\mathbf{p}$. After replacing $\mathbf{\tilde{g}}$ with equation \eqref{eqn:gsol} and $\mathbf{\tilde{p}}$ with equation \eqref{eqn:psol}, we have the following regularization framework,
\begin{equation} \label{eqn:rdreg}
\mathbf{ \Omega(p) = || (1-\beta)A(I-\beta \bar{P})^{-1}p - (1-\alpha)(I-\alpha \bar{G})^{-1}g ||^2 + \kappa ||p||^2},
\end{equation}
where $\mathbf{||p||^2}$ is a 2-norm regularizer and $\kappa$ is a small constant. 
Equation \eqref{eqn:rdreg} takes the standard form of ridge regression, and thus the closed-form solution $\mathbf{p^*}$ can be derived by
\begin{equation} \label{eqn:rdsol}
\mathbf{p^* = (1-\alpha) (\bar{A}^T\bar{A} + \kappa I)^{-1}\bar{A}^T(I-\alpha \bar{G})^{-1}g}.
\end{equation}
where $\mathbf{\bar{A}=(1-\beta)A(I-\beta \bar{P})^{-1}}$. Note that the solution $\mathbf{p^*}$ is a real vector, which can be seen as an approximation of the binary vector $\mathbf{p}$. A simple post-processing is to select one or a few phenotypes that are assigned with significantly larger scores as the phenotypes associated with the gene set. The full algorithm to solve the ridge regression model is given in Fig. \ref{fig:rcNet_alg}. The steps at line 2, 3 and 4 require cubic matrix inversion algorithms. Thus, the time complexity of rcNet algorithm is $\mathbf{O(m^3 + n^3 )}$.


\begin{figure*} 
\centering
\framebox{
  \begin{minipage}{0.45\textwidth}
    $\mathbf{dgsa\_rcNet\_enu(g, \bar{G}, \bar{P}, A,\alpha,\beta)}$
    \begin{enumerate}
    \item[1] $\mathbf{\tilde{g}=(1-\alpha)(I-\alpha \bar{G})^{-1}g}$
    \item[2] $\mathbf{p = 0, s=0}$
    \item[3] $\mathbf{\textrm{{\bf for} }i=1\textrm{ {\bf to} }n}$
    \item[4] ~~~~$\mathbf{p_i=1}$
    \item[5] ~~~~$\mathbf{\tilde{p}=(1-\beta)(I - \beta \bar{P})^{-1} p}$.
    \item[6] ~~~~$\mathbf{s_i=corr(A \tilde{p}, \tilde{g})}$ or $\mathbf{-\sum_{i,j} A_{i,j} (\tilde{p}_i - \tilde{g}_j)^2}$    
    \item[7] ~~~~$\mathbf{p_i=0}$
    \item[8] $\mathbf{j = argmax_i \textrm{ } s_i}$
    \item[9] $\mathbf{p_j=1}$    
    \item[10] \textbf{return} ($\mathbf{p}$)
    \end{enumerate}
  \end{minipage}
}
\caption{$\textrm{rcNet}_{\textrm{corr}}$ and $\textrm{rcNet}_{\textrm{lap}}$ Algorithms - Rank Coherence in Networks by Enumeration.} \label{fig:rcNetEnu_alg}
\end{figure*}

\subsubsection*{Enumeration Methods}
The ridge regression model provides an approximation solution, but if we are only interested in retrieving the most relevant disease phenotype. We can simply go through each phenotype and compute a score against the query gene set $\mathbf{g}$ for each case. Finally, the phenotype with the largest score is chosen as the target phenotype. We propose two functions to measure rcNet for this approach, 
\begin{eqnarray}
\mathbf{rcNet_{corr}(\tilde{g}, \tilde{p}, A)}&=&\mathbf{corr(A \tilde{p}, \tilde{g})},\\
\mathbf{rcNet_{lap}(\tilde{g}, \tilde{p}, A)} &=& - \mathbf{\sum_{i,j} A_{i,j} (\tilde{p}_i - \tilde{g}_j)^2}.
\end{eqnarray}
Function $\textrm{rcNet}_{\textrm{corr}}$   simply uses the Pearson correlation coefficient to check the consistency between $\mathbf{A \tilde{p}}$ and $\mathbf{\tilde{g}}$, similar to the concordance score used by CIPHER \cite{wu:network}. Function $\textrm{rcNet}_{\textrm{lap}}$ checks if the neighboring genes and phenotypes in the association network are assigned similar scores, and the smaller the disagreement, the higher the relevance. This enumeration strategy is similar to CIPHER \cite{wu:network}. The advantages are the conceptual simplicity and the optimality of the exact solution. The disadvantages are the computational cost incurred by the repeated calculation of the association score for each possible combination of the individual phenotypes, and the inflexibility to extend to more general problem of finding multiple target phenotypes. The full algorithm to solve the two enumeration models is given in Fig. \ref{fig:rcNetEnu_alg}. Inside the for-loop between line 3 and 7, the rcNet score is computed for each configuration of $\mathbf{p}$. The overall time complexity of this algorithm is also $\mathbf{O(m^3 + n^3 )}$ if $\mathbf{(1-\beta)(I - \beta \bar{P})^{-1}}$ is precomputed. Note that this is the computational cost by which we only want to retrieve one phenotype. If we want to explore all possible configurations of $\mathbf{p}$, the total cost is exponential in $\mathbf{m}$.


\section*{Results}
The rcNet algorithms are first compared to other methods in experiments of leave-one-out cross-validation and a task of predicting recently discovered disease-gene associations with OMIM data. The rcNet algorithms are then applied to validate findings in datasets from GWAS, DNA copy number analysis, and microarray gene expression profiling.

\subsection*{Preparing Networks}
The disease phenotype network is an undirected graph with 5080 vertices representing OMIM disease phenotypes, and edges with weights in $\mathbf{[0,1]}$. The edge weights measure the similarity between two phenotypes by their overlap in the text and the clinical synopsis in OMIM records, calculated by text mining \cite{van:text}.

The disease-gene associations are represented by an undirected bipartite graph with edges connecting phenotype nodes with their causative gene nodes.  Two versions (May-2007 Version and May-2010 Version) of OMIM associations were used in the experiments \cite{Mc:omim}. The May-2007 Version contains 1393 associations between 1126 disease phenotypes and 916 genes, and the May-2010 Version contains 2469 associations connecting 1786 disease phenotypes and 1636 genes. The May-2007 version was used in the validation experiments on the OMIM data and the GWAS datasets, and the May-2010 version was used in the experiments on the DNA copy number and gene expression datasets. 

Two gene networks were used in the experiments. The first one was derived from the human protein-protein interaction (PPI) network obtained from HPRD \cite{peri03}. The PPI network contains 34,364 binary undirected interactions between 8919 genes. This network was used in the experiments on the OMIM data. A larger human functional linkage network \cite{Huttenhower09} was used in the experiments on the GWAS, DNA copy number and gene expression datasets. This network  contains  24,433 genes and around 60 million weighted edges. To reduce the computational complexity, we applied a cutoff 0.6 on the edge weights to generate a sparser network with around 7 million weighted edges.

\subsection*{Comparison with Other Methods and Evaluations}
The rcNet algorithms were compared with CIPHER \cite{wu:network} and Random Walk with Restart (label propagation) methods \cite{sebastian08, Hwang:heterogeneous, vanunu:assocating, Li:genome-wide}, since those methods reported the best performance for disease gene prioritization. We adopted CIPHER with direct neighbor (C-DN) or shortest path (C-SP) for disease phenotype and gene set association analysis by averaging the correlations across the genes in the query gene set. The Random Walk algorithm described in \cite{Li:genome-wide} (RWR) was chosen as the label propagation method for comparison because it is straightforward to use the model for disease phenotype and gene set association analysis. The two hyper-parameters $\mathbf{\alpha}$ and $\mathbf{\beta}$ for  rcNet were chosen from $\mathbf{\{0.1, 0.5, 0.9\}}$, and a small number $\mathbf{\kappa=10^{-5}}$ was chosen for ridge regression in all experiments. The three balancing parameters 
for RWR were also chosen from $\mathbf{\{0.1, 0.5, 0.9\}}$. For all the methods, the results produced by the best parameters in the leave-one-out cross-validation were reported.

In all the experiments, a query gene set was used to rank all the 5080 disease phenotypes. The higher the target phenotype in the ranking, the better the performance. We measured the performance of a method with receiver operating characteristic (ROC) score, also called area under curve (AUC). Since we are most interested in whether the target phenotype is near the top, we report the the area under the ROC curve up to the first 50 and 100 false positives. 
Another important evaluation is how well a method selects highly coherent top-ranked genes and top-ranked phenotypes since high coherence implies a good utilization of known associations in the model. Specifically, the top genes and phenotypes ranked by the query gene set and the target disease phenotype are assigned largest scores in the cost functions, and connections cancels out the large scores to give smaller penalty. To quantify the connectivity, the top-${\mathbf{r}}$ disease genes and the top-${\mathbf{l}}$ disease phenotypes with known OMIM disease-gene associations are selected to measure \emph{fold enrichment}, which is calculated as $\mathbf{\frac{k}{(r*l)*e}}$, where $\mathbf{k}$ is the number of observed OMIM associations between the $\mathbf{r}$ genes and the $\mathbf{l}$ disease phenotypes, and $\mathbf{e}$ is the probability of observing a random association between a gene node and phenotype node, estimated from the density of the OMIM disease phenotype-gene associations. 
Higher fold enrichment indicates higher coherence between top-ranked genes and disease phenotypes, i.e. highly connected with OMIM disease phenotype-gene associations.


\begin{table}
\centering \caption{ \textbf{Performance comparison in leave-one-out cross-validation and new association prediction with OMIM data.} The tables report the average ROC$_{50}$ and ROC$_{100}$ across all the query cases for each method.
} \label{tab:loo} \scriptsize
\textbf{(A) Leave-one-out cross-validation}\\
\vspace{2pt}
\begin{tabular}{c|c|c|c|c|c|c}
\textbf{Methods} & $\mathbf{rcNet}$ & $\mathbf{rcNet_{corr}}$  & $\mathbf{rcNet_{lap}}$ & $\mathbf{RWR}$ &  $\mathbf{CDN}$ & $\mathbf{CSP}$\\
\hline
\textbf {ROC$\mathbf{_{50}}$} & \textbf{0.160} & \textbf{0.195} & \textbf{0.198} & 0.140 & 0.139 & 0.154\\
\textbf {ROC$\mathbf{_{100}}$} & \textbf{0.206} & \textbf{0.254} & \textbf{0.257} & 0.193 & 0.197 & 0.195\\
\end{tabular}
\vspace{2pt}

\textbf{(B) Prediction of novel disease phenotype-gene associations}\\
\vspace{0pt}
\begin{tabular}{c|c|c|c|c|c|c}
\textbf{Methods} & $\mathbf{rcNet}$ & $\mathbf{rcNet_{corr}}$  & $\mathbf{rcNet_{lap}}$ & $\mathbf{RWR}$ &  $\mathbf{CDN}$ & $\mathbf{CSP}$\\
\hline
\textbf {ROC$\mathbf{_{50}}$} & \textbf{0.117} & \textbf {0.134} & \textbf{0.136} & 0.110 & 0.077 & 0.062\\
\textbf {ROC$\mathbf{_{100}}$} & \textbf{0.151} & \textbf{0.177} & \textbf{0.178} & 0.148 & 0.103 & 0.096\\
\end{tabular}
\label{tab:roc}
\vspace{-10pt}
\end{table}

\begin{figure}[htb]
\begin{center}
\begin{tabular}{c}
\resizebox{!}{3in}{\includegraphics{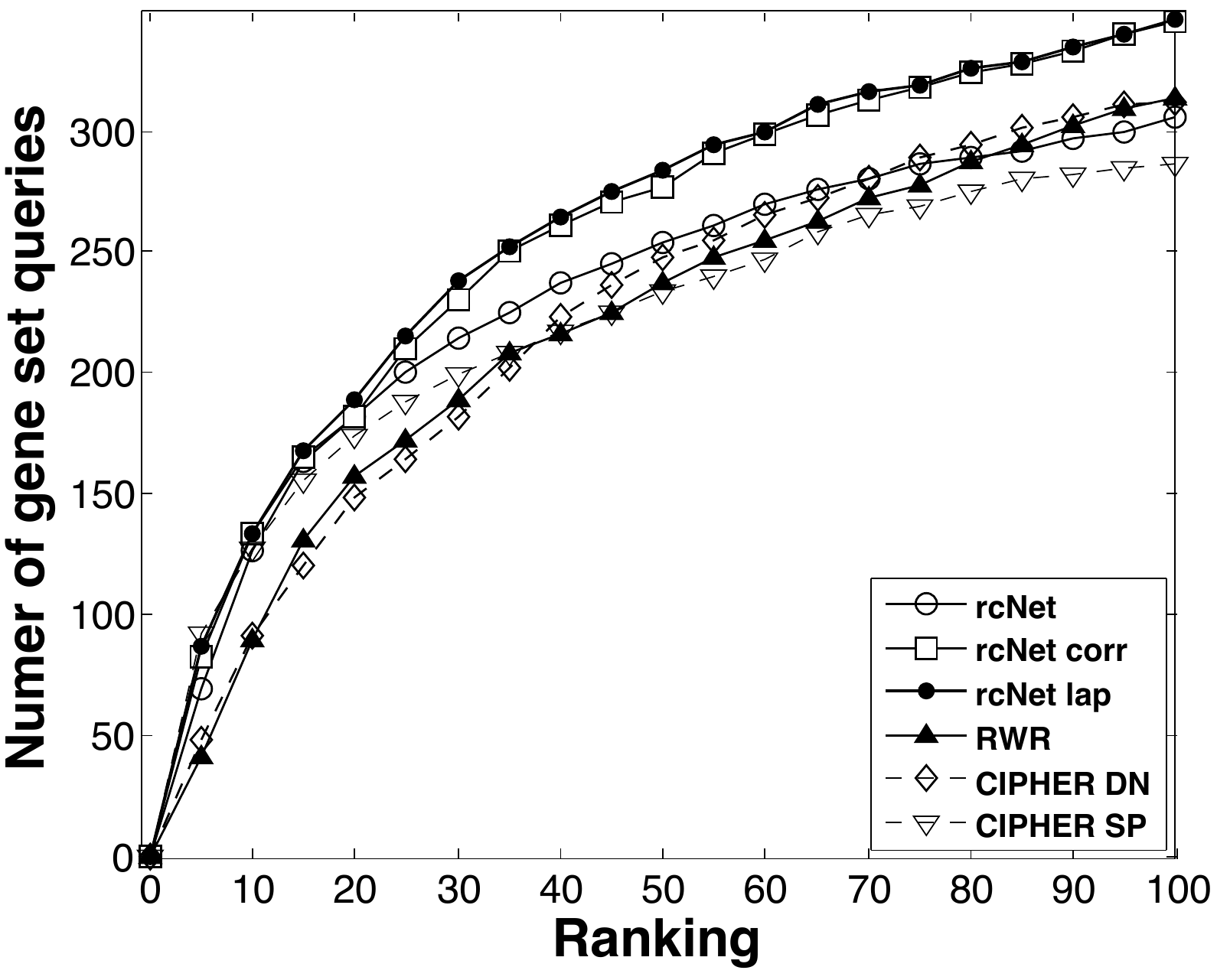}}\\
\end{tabular}
\end{center}
\vspace*{-15pt}
\caption{\footnotesize {\bf Ranking comparison in leave-one-out cross-validation.}
This figure reports the number of query cases, on which a method ranked the target disease phenotype among the top $\mathbf{k \in[1,100]}$ phenotypes. }
\label{fig:roc}
\vspace*{-10pt}
\end{figure}

\subsection*{Leave-one-out Cross-validation in OMIM}
\label{sec:LOO} 
\begin{figure}[htb]
\begin{center}
\begin{tabular}{c}
\resizebox{!}{3in}{\includegraphics{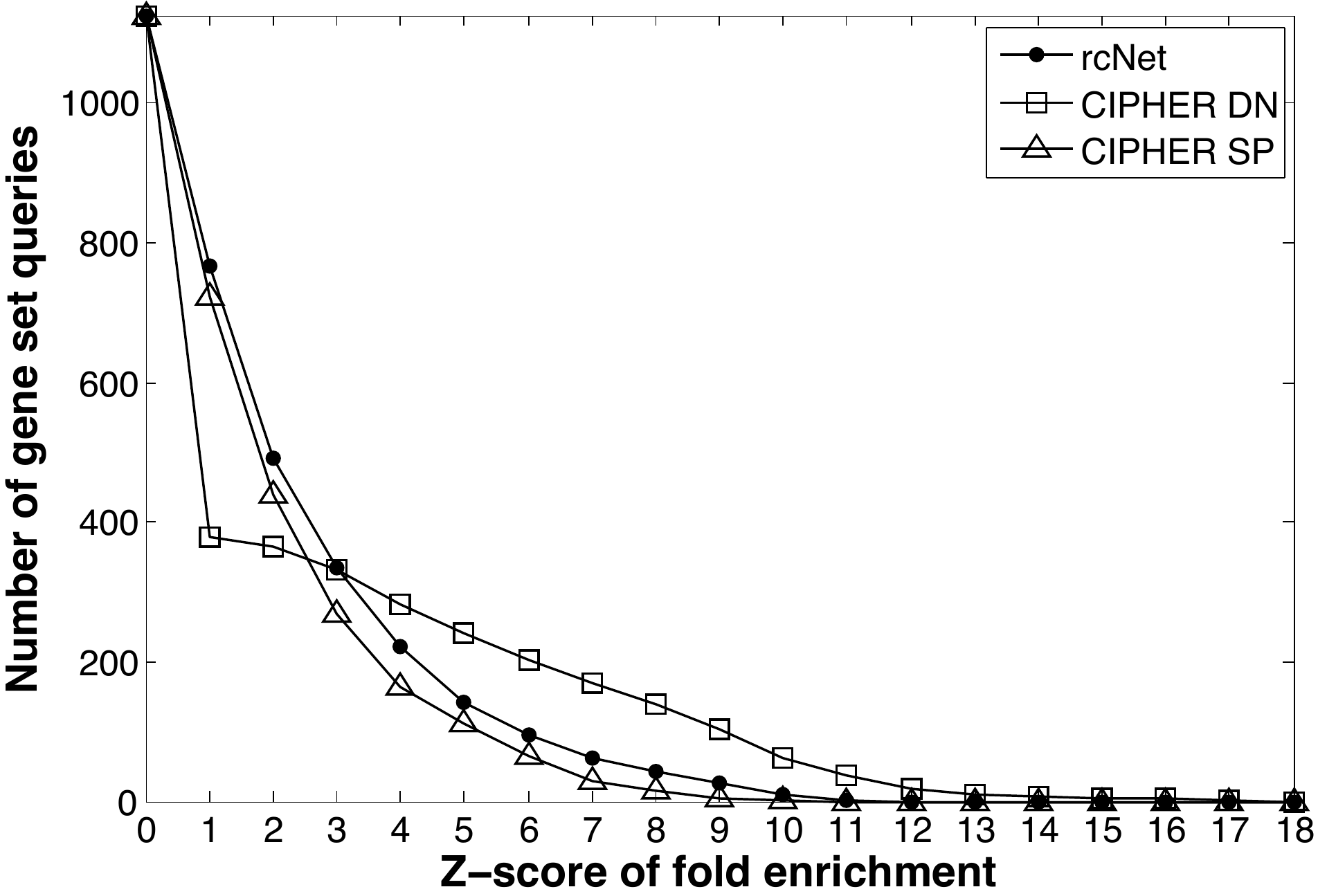}}\\
\end{tabular}
\end{center}
\vspace*{-15pt}
\caption{\footnotesize {\bf Fold enrichment significance.} For each query, the association fold enrichment between the top-20 genes ranked by the query gene set and the top-20 disease phenotypes ranked by each disease phenotype is calculated. A $z$-score of fold enrichment was then computed for the target disease phenotype based on the scores of the unrelated disease phenotypes. This figure plots the number of query cases with a $z$-score above varying thresholds. A $z$-score at 1.96 corresponds to a $p$-value=0.05, assuming a normal distribution.}
\label{fig:fold}
\vspace*{-10pt}
\end{figure}

For each disease phenotype, the genes associated with the phenotype in OMIM were used as the query gene set to retrieve the disease phenotype. Note that the associations between the query gene set and all disease phenotypes including the target disease phenotype were removed in the experiment for leave-one-out cross-validation. In the experiments with RWR, as suggested by \cite{Li:genome-wide}, the disease phenotype network was pruned by taking the 5 nearest neighbors of each node to reduce the computational complexity in leave-one-out cross-validation. Table \ref{tab:roc}(A) reports the average ROC$_{50}$ and ROC${_{100}}$ scores across all the query cases in the leave-one-out cross-validation. Overall,  the rcNet algorithms outperformed the other methods. Specifically, $\textrm{rcNet}_{\textrm{corr}}$ and $\textrm{rcNet}_{\textrm{lap}}$ achieved the best results with about 5$\%$ and 6$\%$ better ranking compared with the best of the others. rcNet performed slightly better than RWR, while CIPHER DN and CIPHER SP achieved lower scores. Fig. \ref{fig:roc} shows a global comparison of the ranking by plotting the number of query cases with the target disease phenotype ranked above a certain rank. Clearly,  the rcNet algorithms achieved better rankings at any ranking threshold in the experiments. For example, $\textrm{rcNet}_{\textrm{corr}}$ and $\textrm{rcNet}_{\textrm{lap}}$ ranked around 290 query cases above rank 50, while RWR and CIPHERs ranked around 230 query cases above the rank. \\ 
We further analyzed how the rank coherence between the rankings by the query gene set and the disease phenotypes could affect the performance of rcNet algorithms and CIPHER. Based on the coherence assumption, the top ranked genes and the top ranked phenotypes by the query gene set and the target disease should be highly connected with each other by a good method. Fig. \ref{fig:fold} compares the number of queries that achieved a significant fold enrichment for the target disease phenotype compared with the unrelated phenotypes. rcNet consistently identified more cases with significant fold enrichment against CIPHER SP at all the $z$-score thresholds. This observation suggests that label propagation is a better measure than shortest path to distinguish a target phenotype from unrelated ones because the information of gene neighborhoods are better utilized. Interestingly, CIPHER DN detected much less associations with high significance, but more cases with very high significance. For some query gene sets, which include genes with many disease genes as direct neighbors in dense disease gene modules, the direct gene neighbors tend to have more dense associations with the related phenotypes. However, only less than one-third of the query gene sets are the easy cases. CIPHER DN failed to find a significant fold enrichment for the other two-thirds. Thus, CIPHER DN is not performing well in general. 


\subsection*{Predicting new OMIM Associations}
\label{sec:OMIM} 
To further evaluate how well a method can predict new disease-gene associations based on known associations, a case study was designed to predict the target disease phenotype of the new disease genes added into OMIM between May, 2007 and May, 2010. There are 387 new disease phenotypes were annotated in OMIM since May, 2007, excluding 11 new disease phenotypes whose disease genes have no interaction in the gene network. In this experiment, the task is to predict the target disease phenotype of the newly annotated disease genes, i.e. to query a set of new disease genes of a disease phenotype to retrieve the phenotype based on the disease-gene associations in May-2007 Version. Table \ref{tab:roc}(B) reports the average ROC$_{50}$ and the ROC${_{100}}$ scores. $\textrm{rcNet}_{\textrm{corr}}$ and $\textrm{rcNet}_{\textrm{lap}}$ performed the best, followed by rcNet and RWR, and CIPHER DN and CIPHER SP did not produce comparable results with the other methods. A possible reason for the worse performance of CIPHER is that the new cases are relatively under studied compared with the other disease phenotypes, and the global information in all the networks are necessary for an accurate inference of the associations. The results further supports the better performance of the rcNet algorithms compared with the baselines. 

\begin{table*}[!htb]
 \caption{{\bf Ranking the target disease phenotype of the disease susceptibility genes identified from GWAS.} The disease categories in the first column are based on the definition in \cite{goh:network}. In the third column, the PubMed IDs marked with `*' denote multiple GWASs for a disease/trait. Refer to supplementary Table for the results of the full list of the GWAS cases.}\label{tab:casestudy_GWAS}
\scriptsize
\centering

\begin{tabular}{|c|c|c|c|c|c|c|}
\hline
 \multirow{2}{*}{\bf Category} & \multirow{2}{*}{\bf Disease/Trait}	  &	\multirow{2}{*}{\bf OMIM Index}	&	{\bf Gene Set}	&	{\bf Rank by}	&	{\bf Rank by}	&	{\bf Rank by} 	\\	
 & & &  {\bf Size} & {\bf rcNet} & $\mathbf{rcNet_{corr}}$ & $\mathbf{rcNet_{lap}}$ \\	\hline


\multirow{9}{*}{Cancer}	&	Prostate cancer		&	176807	&	15	&	2	(0.03\%)	&	2	(0.03\%)	&	2	(0.03\%)	\\	
	&	Breast cancer		&	113705	&	26	&	7	(0.1\%)	&	51	(1\%)	&	43	(0.8\%)	\\	
	&	Basal cell carcinoma (cutaneous)	&		605462	&	5	&	7	(0.1\%)	&	189	(3.7\%)	&	228	(4.5\%)	\\	
	&	Basal cell carcinoma (cutaneous)	&		604451	&	5	&	90	(2\%)	&	202	(4\%)	&	256	(5\%)	\\	
	&	Urinary bladder cancer	&		109800	&	1	&	14	(0.2\%)	&	48	(0.9\%)	&	60	(1.1\%)	\\	
	&	Acute lymphoblastic leukemia (childhood)	&		159555	&	3	&	19	(0.04\%)	&	51	(1.0\%)	&	45	(0.8\%)	\\
	&	Lung cancer	&	211980	&	12	&	22	(0.4\%)	&	587	(12\%)	&	1610	 (32\%)	\\	
	&	Lung adenocarcinoma	&	211980	&	6	&	52	(1\%)	&	838	(16\%)	&	1815	 (36\%)	\\	
	&	Chronic lymphocytic leukemia	&	151430	&	14	&	57	(1\%)	&	318	(6.3\%)	&	306	(6\%)	\\	
	&	Neuroblastoma (high-risk)	&	600613	&	1	&	143	(3\%)	&	110	(2\%)	&	138	(3\%)	\\	\hline
\multirow{3}{*}{Immunological}	&	Systemic lupus erythematosus	&	152700	&	10	&	46	(0.9\%)	&	178	(4\%)	&	161	(3\%)	\\	
	&	Leprosy	&	246300	&	4	&	78	(1.5\%)	&	62	(1.2\%)	&	64	(1.3\%)	\\	
	&	Leprosy	&	607572	&	4	&	272	(5\%)	&	54	(1\%)	&	55	(1\%)	\\	\hline
\multirow{2}{*}{Endocrine}	&	Type 2 diabetes	&	125853	&	9	&	97	(2\%)	&	718	(14\%)	&	1912	 (38\%)	\\	
	&	Type 1 diabetes	&	222100	&	26	&	331	(7\%)	&	690	(13\%)	&	191	(3.8\%)	\\	\hline
Gastrointestinal	&	Crohns disease	&	266600	&	2	&	60	(1.2\%)	&	1396	 (27\%)	&	3012	 (59\%)	\\	\hline

\end{tabular}
\label{tab:gwas}
\end{table*}

\subsection*{Predicting Disease Phenotypes of Disease Susceptibility Genes from GWAS}
The goal of GWAS is to discover disease susceptibility loci/genes that could be useful for assessing or predicting an individual's risk of disease. However, it is often challenging to assess how a set of novel disease susceptibility genes potentially influence susceptibility in disease, especially when the set of genes have no or little previously known disease implications, or function and pathway annotations. 
In this case study, we collected new disease susceptibility genes from GWAS, whose roles in disease susceptibility are not previously understood, and applied rcNet algorithms to predict the disease phenotype of the disease susceptibility genes. We extracted all the disease susceptibility genes discovered in GWAS based on a recent survey of all studies reported in the GWAS catalog as of Dec. 2010 \cite{hindorff2009}.
After filtering out the genes already included in OMIM May-2007 Version, we selected 217 diseases/traits with novel susceptibility genes that are not associated with any disease phenotype in OMIM May-2007 Version, and 31 out of the 217 diseases/traits could be matched with OMIM phenotypes in the disease network. Subsequently, the 31 diseases/traits and their susceptibility genes were used in this experiment. 

\begin{figure*}
  \centering
  \includegraphics[scale=0.6]{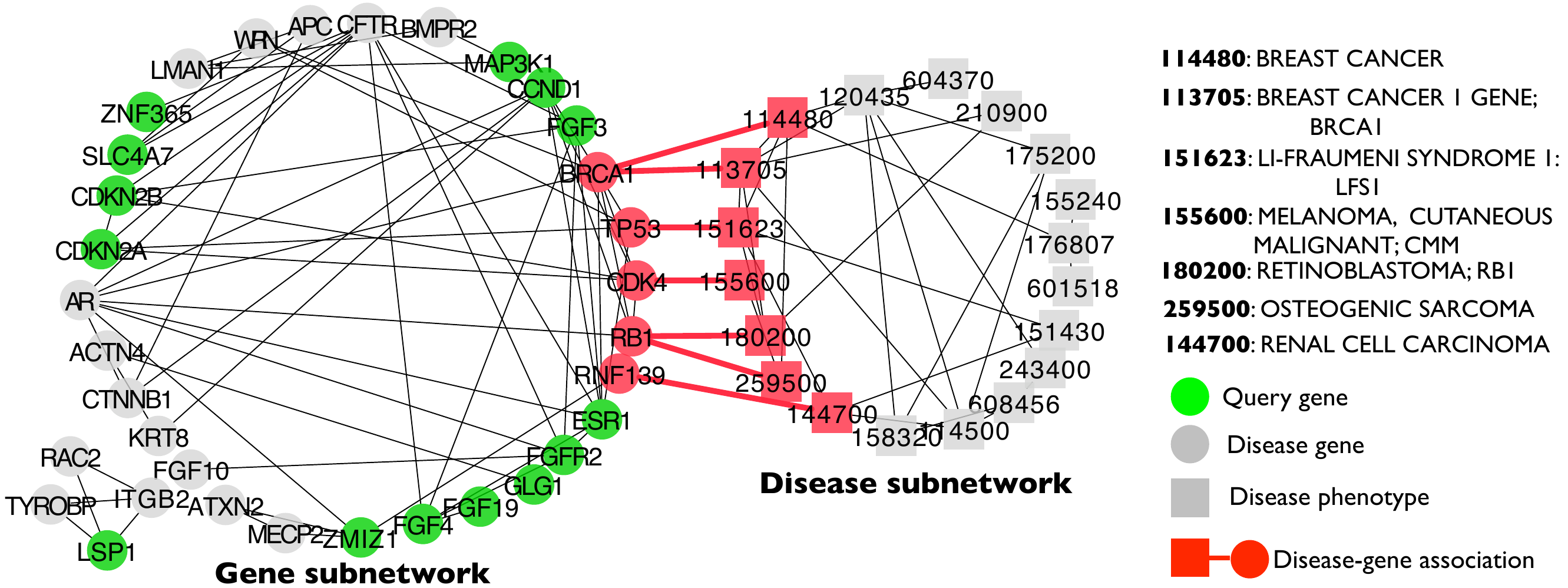}
  \vspace{-10pt}
\caption{{\bf Querying with breast cancer susceptibility genes from GWAS by rcNet.} This example shows how rcNet algorithm  predicted the target disease phenotypes of breast cancer susceptibility genes from GWAS. By querying with the 26 novel breast cancer susceptibility genes from GWAS, rcNet ranked the 20 disease genes in the gene subnetwork at the top. The gene subnetwork also includes 14 out the 26 query genes, which are connected with the top-20 genes. Similarly, the top 20  disease phenotypes ranked by OMIM114480:breast cancer disease phenotype are included in the disease subnetwork. 
In this example, 5 of the 20 top-ranked disease genes are connected to 7 of the top-20 disease phenotypes given by 7 OMIM disease-gene associations, compared with the expected 0.87 association between 34 random genes and 20 random phenotypes.} \label{fig:gwas}
  \vspace{-10pt}
\end{figure*}

We queried the set of disease susceptibility genes of each of the 31 diseases/traits to rank the 5080 OMIM disease phenotypes. A disease/trait could be matched with multiple OMIM disease phenotypes. 
We report the rank of the matched phenotype with the best ranking for each query. The ranking results of a subset of the 31 diseases/traits are reported in table \ref{tab:gwas}. 
Among the 31 queries, 14 cases ranked the target diseases within top 2$\%$ (ranked within top 100).  Notable examples are prostate cancer, breast cancer, basal cell carcinoma, bladder cancer,  acute lymphoblastic leukemia, systemic lupus erythematous, and leprosy. In these cases, the rcNet algorithms ranked the target disease phenotype of the query gene set within top  1$\%$. Fig. \ref{fig:gwas} shows the example that rcNet accurately ranked the breast cancer phenotypes of the breast cancer susceptibility genes, by querying with 26 novel breast cancer susceptibility genes from GWAS. The connectivity between the top ranked disease genes and the top ranked disease phenotypes is around 13 folds of the expected number of connections between the same numbers of random genes and phenotypes, or around 11 folds of the average number of connections given from the phenotype ranking by the relevance to the unrelated disease phenotypes. One interesting observation is that the target disease phenotype OMIM:113705 `BREAST CANCER 1 GENE; BRCA1' is only directly connected with two top ranked disease phenotypes, OMIM: 114480 `BREAST CANCER', OMIM:151623 `LI-FRAUMENI SYNDROME 1:LFS1', and OMIM:`259500: OSTEOGENIC SACROMA', and only 5 of the 26 query genes directly interact with the top ranked disease genes with disease-gene associations. The neighbor expansions in both the gene network and the phenotype network resulted in 4 OMIM disease-gene associations. This observation suggests that, simply exploring the direct neighbors of the query gene set and the target disease phenotype in the networks, a method might fail to infer disease-gene set associations, due to the low statistical significance of the sparse connectivity between the genes and the disease phenotypes. Specifically, in this example, the fold enrichment for 4 associations is 7.53, which is significantly lower than the 12.35 fold enrichment obtained by rcNet. 
Another interesting example is the inference of the association between leprosy and its susceptibility genes from GWAS (pubmed 20018961). In OMIM May-2007 Version, leprosy has no causative genes, and the leprosy susceptibility genes from GWAS also have no association with any disease. The lack of known associations in both the target disease phenotype and the get set poses a hard case that gene set enrichment analysis based on overrepresentation will fail to reveal, but rcNet algorithms ranked Leprosy within top 2$\%$.

In contrast to the results in cross-validation on OMIM data, rcNet produced significantly better results in cancer, immunological, and gastrointestinal disease, compared with $\textrm{rcNet}_{\textrm{corr}}$ and $\textrm{rcNet}_{\textrm{lap}}$. Interestingly, previous studies showed that disease susceptibility genes from GWAS catalog have less modularities in the gene network compared to those of the known disease genes in OMIM, and phenotypically similar diseases such as immunological and gastrointestinal diseases do not tend to share their disease genes \cite{Baranzini09, Barrenas09}. Those previous studies also implicated that due to the unique topological characteristics of disease susceptibility genes discovered in GWAS, the existing network-based methods would fail to reveal the associations between the disease susceptibility genes and the disease. However, our experiments suggest that, by incorporating the global topological information in the networks and the known OMIM associations, rcNet algorithms can successfully discover the elusive associations in many cases.  

\subsection*{Predicting Disease Phenotypes of Genes with Copy Number Changes} \label{sec:CNV}
In DNA copy number analysis, genes in the chromosomal regions with copy number changes are identified as candidate disease genes. In this experiment, we applied the rcNet algorithms to predict the target disease phenotypes of the candidate disease genes in disease susceptible copy number change regions. We collected 13 human DNA copy number change datasets from a recent human cancer copy number study from \url{http://www.broadinstitute.org/tumorscape} \cite{Beroukhim10}. The DNA copy number measurements in the datasets were obtained on the Affymetrix 250K Sty SNP array. 
The regions with copy number changes were detected by GISTIC tool with default settings \cite{GISTIC}. Genes in the detected copy number change regions were used as the query gene set to predict their target disease phenotypes. 
Table  \ref{tab:cnv} shows the ranking results by the rcNet algorithms. rcNet ranked the target disease within top 2$\%$ for 6 of the 13 cancers and $\textrm{rcNet}_{\textrm{corr}}$ ranked the target disease  within top 2$\%$ for 7 of the 13 cancers. In 9 of the cases, at least one algorithm ranked the target disease within top 100. \cite{Beroukhim10} stated that more than three-quarters of the statistically significantly altered copy number regions contain potential cancer causing genes that are not previously validated targets of somatic copy number alternations in human cancer. This suggests that enrichment analysis of the genes will not reveal any disease-association, but rcNet algorithms found many associations with the network information.   

\begin{table}[!htb]
 \caption{{\bf Ranking the target disease phenotypes of the candidate disease genes with copy number changes.} This experiment includes 13 human cancer copy number studies from \cite{Beroukhim10}.}\label{tab:casestudy_CNV}
\scriptsize
\centering
\begin{tabular}{|c|c|c|c|}
\hline
\multirow{2}{*}{\bf Disease/Trait} &	{\bf Rank by}	&	{\bf Rank by}	&	{\bf Rank by} 	\\	
&  {\bf rcNet} & $\mathbf{rcNet_{corr}}$ & $\mathbf{rcNet_{lap}}$  \\	\hline
Neuroblastoma	&	5	&	13	&	126	\\	\hline
Colorectal cancer	&	14	&	20	&	613	\\	\hline
Renal cancer	&	22	&	14	&	33	\\	\hline
Non small cell lung	cancer &	34	&	48	&	558	\\	\hline
Breast cancer	&	68	&	136	&	521	\\	\hline
Medulloblastoma	&	77	&	826	&	2007	\\	\hline
Prostate cancer	&	129	&	127	&	2447	\\	\hline
Ovarian cancer	&	322	&	73	&	1108	\\	\hline
Small cell lung cancer	&	759	&	53	&	909	\\	\hline
Mesothelioma	&	959	&	21	&	54	\\	\hline
Gastrointestinal stromal tumor	&	1169	&	787	&	1679	\\	\hline
Hepatocellular carcinoma	&	4241	&	952	&	1295	\\	\hline
Glioma	&	4705	&	787	&	951	\\	\hline
\end{tabular}
\label{tab:cnv}
\end{table}

\subsection*{Predicting Disease Phenotypes of Differentially Expressed Genes}
\label{sec:GE}
It is frequently observed that many disease susceptibility genes are not differently expressed in microarray gene expression experiments. 
In this experiment, we applied rcNet algorithms to predict the target disease of differentially expressed genes in gene expression profiles.  We collected 13 human cancer microarray gene expression dataset from GEO. Gene expression profiles were obtained on the Affymetrix HU133A array, and normalized by RMA \cite{irizarry2003ena}. Standard $t$-test was used to identify differentially expressed genes. The differentially expressed genes were used to query for their target diseases. To quantify how reliable a differential expression is,  the query gene nodes were initialized by the absolute values of the  $t$-statistics for label propagation.  Table \ref{tab:ge} reports the results of predicting the target diseases of the differentially expressed genes. Out of the 13 cases, the rcNet algorithms  could rank 7 within top 5$\%$, and 12 within top 10\%. Although the result is only moderately encouraging, it validates the hypothesis that the neighboring information of the differentially expressed genes provides clue of association with the target disease phenotype.



\begin{table}[!htb]
 \caption{{\bf Ranking the target disease of differentially expressed genes.} The first column represents the target disease of a microarray gene expression study, and the second column gives the GEO number of the dataset. }\label{tab:casestudy_GE}
\scriptsize
\centering
\begin{tabular}{|c|c|c|c|c|}
\hline
\multirow{2}{*}{\bf Disease/Trait}& {\bf GEO} &	{\bf Rank by}	&	{\bf Rank by}	&	{\bf Rank by} 	\\	
&  {\bf Num.}& {\bf rcNet} & $\mathbf{rcNet_{corr}}$ & $\mathbf{rcNet_{lap}}$  \\	\hline
AML	&	GSE9476	&	576	&	316	&	359	\\	\hline
\multirow{5}{*}{Breast cancer}	&	GSE7390	&	14	&	49	&	51	\\	
	&	GSE2034	&	40	&	130	&	146	\\	
	&	GSE6532	&	129	&	151	&	182	\\	
	&	GSE1456	&	138	&	102	&	109	\\	
	&	GSE3494	&	161	&	709	&	1313	\\	\hline
Gastric cancer	&	GSE13911	&	248	&	298	&	362	\\	\hline
\multirow{3}{*}{Lung cancer}	&	GSE10072	&	206	&	755	&	2219	\\	
	&	E-MEXP-231	&	318	&	608	&	1115	\\	
	&	GSE7670	&	379	&	1330	&	4002	\\	\hline
Ovarian cancer	&	GSE6008	&	414	&	1494	&	2283	\\	\hline
\multirow{2}{*}{Prostate cancer}	&	E-MEXP-1327	&	271	&	1446	&	2057	\\	
	&	GSE8218	&	900	&	1214	&	2498	\\	\hline
\end{tabular}
\label{tab:ge}
\end{table}

\section*{rcNet WebTool}
\begin{figure}[htb]
\begin{center}
\begin{tabular}{c}
\resizebox{!}{3.0in}{\includegraphics{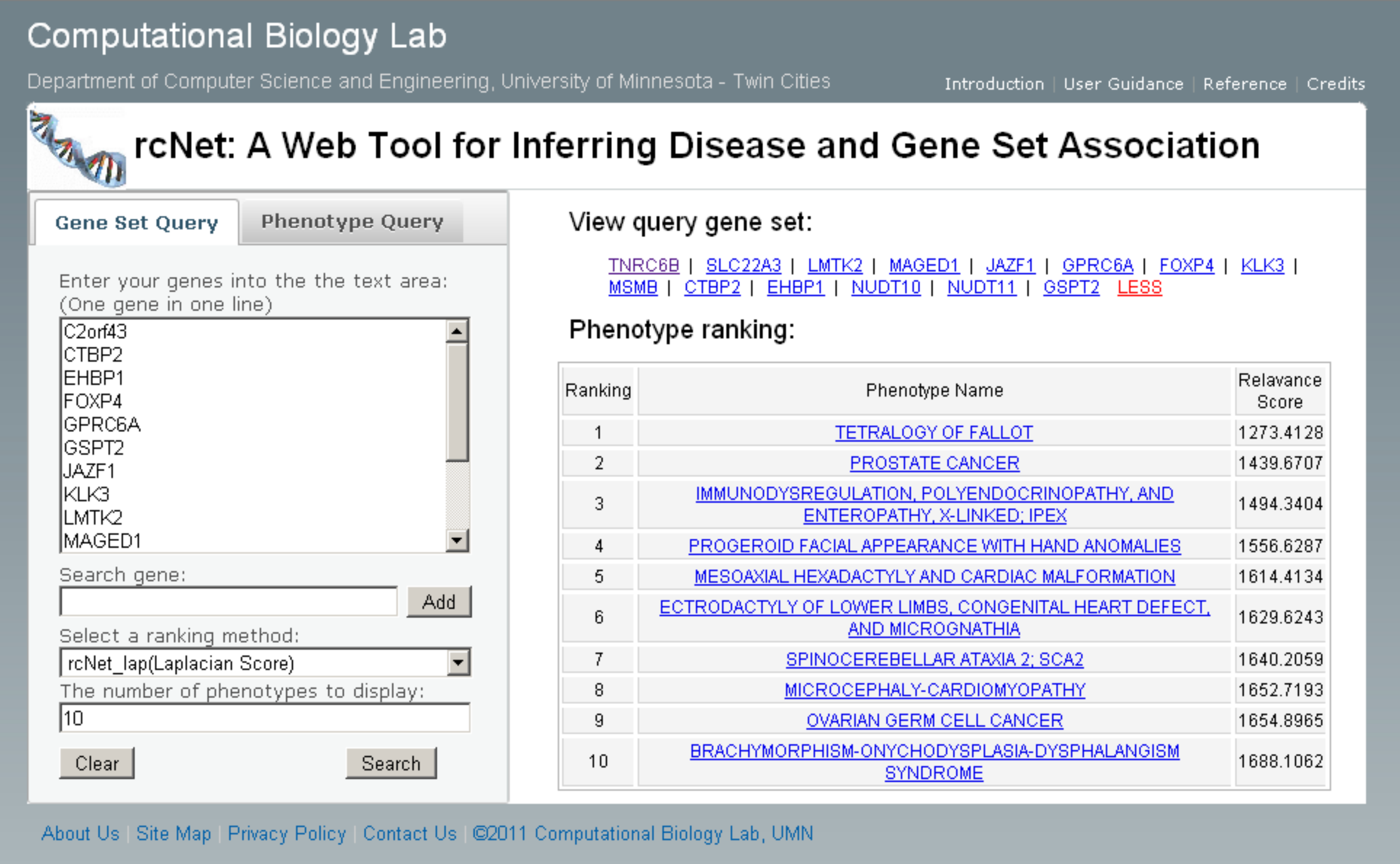}}\\
\end{tabular}
\end{center}
\vspace*{-15pt}
\caption{\footnotesize {\bf rcNet WebTool Demo.} In this example, a gene set with a list of 15 genes identified as prostate cancer susceptibility genes in GWAS was used to query rcNet WebTool. The left panel shows the settings used for query and the right panel displays the query result.}
\label{fig:web}
\vspace*{-10pt}
\end{figure}

The rcNet algorithms were implemented and deployed as a general webtool for disease-gene set association analysis at \url{http://compbio.cs.umn.edu/dgsa_rcNet}. Providing a list of query genes, a user can retrieve the OMIM disease phenotypes ranked by their degree of association with the gene set. In Fig. \ref{fig:web}, we show an example of querying rcNet WebTool with the disease gene set of prostate cancer from GWAS. In the implementation, the Laplacian scores are precomputed to improve efficiency. Currently, it takes the server less than 5 seconds to response to a gene set query. 

\section*{Discussion}
Analysis of the gene sets from genome-wide high-throughput screening is a continuing challenge in many disease studies. When the gene set is poorly annotated, enrichment analysis will fail to detect any associations with disease phenotypes, or when the gene set contains genes in a broad range of functional categories, enrichment analysis provides unreliable statistical significance. Statistics from OMIM (Jan 2011) show that 3745 of the 6675 disease phenotypes are still unknown for their molecular basis. Thus, enrichment analysis will fail to find any associations between the 3745 disease phenotypes and any query gene set. 
For example, in the experiments with the GWAS gene sets, rcNets  algorithms ranked leprosy (OMIM:246300 and OMIM:607572) among the top 2$\%$ phenotypes, while enrichment analysis reported no association for the four disease susceptibility genes of leprosy. 
rcNet focuses on improving detection of disease phenotype-gene set associations by integrating gene network and disease network to better summarize sparse associations for a global comparison of all possible disease and gene set associations. 
The rcNet algorithms effectively utilizes hidden information in the gene network and the disease network with the machine learning models. First, the label propagation steps on both the gene network and the disease network fully explore the neighborhood information of the query genes and a disease phenotype. The relevance information is propagated from the seed nodes to their neighbors to provide a global quantification of relevance, and the relevance scores are then utilized with all the known associations for evaluating the association between the gene set and the disease phenotype. Thus, analysis with rcNet is not biased by poor known annotation or the size of the query gene set. Second, compared with the other methods that also utilizing the gene network and the disease network, rcNet is more flexible in handling the network data because rcNet is capable of handling weighted associations and weighted edges in the gene network and the disease network. rcNet does not rely on deciding direct neighbors or shortest path as CIHPER or PRINCE \cite{vanunu:assocating}. Finally, the ridge regression model coupled with label propagation provides an approximation of finding association between a gene set and multiple disease phenotypes, which is difficult to achieve with enumeration-based strategies. 

Despite the encouraging results of rcNet, there are also limitations. First, rcNet relies heavily on the networks. For the cases where the gene set already has known associations with the target disease phenotype, the network information might introduce noise to dilute the strong signal as showed in Fig. \ref{fig:fold}. Thus, rcNet is more useful for studying new diseases that have not been genetically characterized rather than confirming well-understood diseases. Second, it is also difficult to distinguish the closely related phenotypes from false positives, because it is possible that some of the top-ranked phenotypes is not similar to or share any common disease genes with the target disease phenotype in the disease network. Interpretation of these phenotypes will not be straightforward. A possible solution is to identify subnetworks as the example in Fig. \ref{fig:gwas} and use information from the gene cluster and the phenotype cluster for finding explanations.


rcNet is a helpful tool for oncologists to analyze disease and gene set association. Researchers can validate their findings from high-throughput studies, especially, to validate novel associations between complex diseases and a query gene set with no known associations. rcNet can also help identify closely related phenotypes of the target disease of the query gene set. Since rcNet algorithms utilize both the disease similarities and the gene interactions, some phenotypically similar disease phenotypes will be ranked at the top. These disease phenotypes might provide additional information to investigate the target disease in the study.
In future work, we plan to extend rcNet as a more general tool that could also infer enriched functions of a query gene set. We can build a GO function network with GO term nodes and edges weighted by the similarity between two GO functions. The application of the rcNet algorithm under this context will be straightforward.

\section*{Acknowledgments}
This work is partially supported by IHI Research Seed Grant from the Institute of Health Informatics at University of Minnesota TC.

\bibliographystyle{plos2009}
\bibliography{plos_template}

\end{document}